\documentclass[a4paper, times, 10pt, twocolumn, twoside,top=3.5cm,bottom=3.5cm,left=2.5cm,right=2cm]{article}
\usepackage{multirow}
\usepackage{url}

\usepackage{tikz}
\def\checkmark{\tikz\fill[scale=0.4](0,.35) -- (.25,0) -- (1,.7) -- (.25,.15) -- cycle;} 
\usepackage{breakurl}
\usepackage[breaklinks]{hyperref}

\usepackage{ISARC}

\usepackage{lscape}
\usepackage{hologo}

\usepackage[utf8]{inputenc} 
\usepackage[OT2,T1]{fontenc}    
\usepackage[russian, english]{babel} 
\usepackage{ISARC}
\usepackage{amsmath,amssymb,amsfonts}
\usepackage{algorithmic}
\usepackage{graphicx}
\usepackage{textcomp}
\usepackage{xcolor}

\def\BibTeX{{\rm B\kern-.05em{\sc i\kern-.025em b}\kern-.08em
    T\kern-.1667em\lower.7ex\hbox{E}\kern-.125emX}}

\begin{document}
\linespread{0.5}
\title{Survey: Vitals Screening Techniques for a Safer Environment}

\author{Sarah Baig, Mohammed Elbadry}
\affiliation{
Research and Development Department, Soter Technologies
}

\email{
\href{mailto:sarah.baig@sotertechnologies.com}{sarah.baig@sotertechnologies.com}, 
\href{mailto:mohammed.elbadry@sotertechnologies.com}{mohammed.elbadry@sotertechnologies.com}
}
\maketitle

\begin{abstract}
With COVID-19 disrupting operations across various sectors of the workforce (e.g., offices, airports, libraries, schools), preventative measures enabling resumption of work are quickly becoming a necessity. In this paper, we present the need for Vitals Screening Techniques (VIST) where more than one vital is screened to ensure the  safety of the population (e.g. temperature, heart rate, and blood oxygen levels). VIST can be deployed in crowded environments to provide the new  necessary layer of safety. We provide extensive coverage of state-of-art technology that can assist in tackling this emerging problem, and evaluate one of the existing products on the market that employ VIST. 
\end{abstract}

\section{Introduction}

The COVID-19 Pandemic has fundamentally changed society’s perspective~\cite{hughes2010origin} on common spaces (e.g., airports, schools, commercial establishments). The 'new normal` has heralded a new age of technological innovations, from contact tracing, telemedicine to newer screening methods\cite{barnes2020challenges}. Although temperature screening has been widely implemented \cite{bwire2020coronavirus,aw2020non} as a way to safely reopen, multi-vital screening provides greater insight into the health of an individual. Vital signs, including temperature, blood oxygen levels and heart rate, are the easiest and most critical data points gathered from an individual to assess their general health.~\cite{chen2020cardiovascular}.

In emergency settings, patients have to be prioritized and guided to the correct place of treatment (``triage'') largely based on their vital signs~\cite{barfod2012abnormal, christ2016emergency, australasian2000guidelines,mchugh2012more}. Vitals screening methods are applied to high throughput areas (e.g., airports, businesses, warehouses, factories), especially in the case of a global pandemic~\cite{fda2020enforcement}. One of the vitals that is important during screening is heart rate. Elevated heart rate (Tachycardia) is a telltale sign associated with fever in the case of viral respiratory infections~\cite{karjalainen1986fever}. Another vital that can be helpful is blood oxygen levels (SpO2). Low blood oxygen levels were seen before the onset of a fever in many COVID-19 patients~\cite{chandra2020silent} and can indicate underlying signs of other viral diseases~\cite{chughtai2017presence}. Body temperature, heart rate, and blood oxygen levels together provide further insight into the health of an individual. 

Human body temperature is well established as one of the key vital signs~\cite{geneva2019normal}. The accepted mean value for normal human body temperature measured orally is 37°C (98.6°F). However newer research indicates that the average may actually be closer to 36.6°C  (97.9°F)~\cite{protsiv2020decreasing}. Each individual has their normal body temperature, which varies slightly from the ideal value. Human body temperature constantly adapts to its environmental conditions. A body temperature of 38°C (100.4°F) or more is considered to be a fever~\cite{BT}. The most recent viral epidemics have had fever as the most common symptom (e.g., Ebola, SARS, H1N1~\cite{ghassemi2018best}, and COVID-19~\cite{grant2020prevalence}). It is evident that fever detection is one of the key components in screening. 

 All current screening techniques rely on remote body temperature measurements. There are two problems that must be acknowledged when measuring the body temperature: i) normal body temperature variation; ii) infrared thermal imaging limitations. Body temperature can fluctuate based on the region selected for measurement~\cite{lenhardt2006estimation,lahiri2012medical}. Furthermore, research has shown that body temperature is a nonlinear function of several variables such as age, state of health, gender, environmental temperature, time of the diurnal cycle, and among many others~\cite{geneva2019normal}. On average, healthy elderly people have lower body temperature compared to younger adults~\cite{geneva2019normal}. The human body is constantly adapting its temperature to environmental conditions (e.g., goes up in the afternoon and lower at night). Despite these minor variations, elevated body temperature is still a universally accepted indicator of fever.

Remote body temperature screening is a fast, non-invasive alternative to conventional clinical thermometers for monitoring body temperature~\cite{nguyen2010comparison, lahiri2012medical}. Average external body temperature (peripheral skin temperature) is 2-4°C (3.6-7.2°F) less than the core temperature~\cite{lenhardt2006estimation}. Therefore, mean body temperature must be calculated from external (or skin) temperature using an estimation algorithm. Infrared radiation emitted by a surface depends on the environmental conditions such as moisture, airflow and surrounding temperature~\cite{IPVM,chen2020investigation}. Other factors that impact temperature sensing are ambient temperature drift and aging of the sensor. An individual’s thermal state also affects the radiated heat (e.g., running, coming from cold environments, etc). Further, the distance and angle of the thermal camera relative to the subject plays a critical role in the sensor's fidelity. Blackbody devices (temperature references) are known to solve the issues related to ambient temperature and sensor aging, improving the accuracy of the sensor. However, they are often forgone due to the cost or complexity of deployment~\cite{FLIR}. Remote body temperature sensing is an ideal alternative to clinical thermometers that are sometimes cumbersome and often require an attendant~\cite{nguyen2010comparison, FDA6}.     

Beyond temperature sensing, blood oxygen levels are used to infer any impairment in lung function~\cite{tremper1989pulse}. Blood oxygen level is usually measured with a pulse oximeter (finger clip). A resting oxygen saturation level between 95\% and 100\% is regarded as normal for a healthy person at sea level, and below 95\% is considered abnormal~\cite{Hypoxemia,PulseOximetry, torp2020pulse,hafen2018oxygen}. Low blood oxygen can serve as an indicator to many different viral pneumonias~\cite{chandra2020silent}. The recent global pandemic (i.e. COVID-19) has demonstrated that many people can have dangerously low oxygen levels, without showing any other symptoms ('silent hypoxemia') ~\cite{teo2020early}. The detection of low oxygen levels in asymptomatic individuals can facilitate early diagnosis of an underlying illness. Blood oxygen measurement serves as a key component of Vital Screening. 

Heart rate is measured through pulse oximeters, in addition to blood oxygen levels. The American Heart Association defines the normal sinus heart rate as between 60 and 100 bpm at rest~\cite{avram2019real,d2015crosstalk,karjalainen1986fever} (it is important to note that athletes often have heart rates below 60 bpm at rest). Tachycardia is observed in case of anemia, intake of caffeine, and exercise \cite{Tachycardia}. Tachycardia is seen concomitantly with fever due to an increase in the Basal Metabolic Rate and cardiac output. In one study, when the temperature rose by 1°C (1.8°F) due to fever, the heart rate increased on the average by 8.5 beats per minute~\cite{karjalainen1986fever}. Thus, tachycardia, when seen along with fever, can point to possible infection. 

Pulse oximetry technology involves shining light at specific wavelengths through tissue (most commonly the fingernail bed) and using a detector to determine the amount of light that passes through~\cite{demeulenaere2007pulse, castaneda2018review}. There are several inherent limitations to pulse oximetry. One of the common examples of interfering factors is poor signal due to certain nail polish and artificial fingernails~\cite{luks2020pulse}. Poor peripheral perfusion because of cold, hypotension, or Raynaud's disease is the principal cause for failure to obtain a satisfactory signal, mainly because of an inadequate pulse wave~\cite{torp2020pulse,demeulenaere2007pulse}. Motion artifacts can interfere with signal detection because of an unstable waveform. Improperly seated sensors, shivering, seizures, or tremors can cause movement leading to inaccurate readings. Pulse oximetry, despite its limitations, is universally recognized as an essential vital measurement tool.  

The need to take preventative measures to prepare for inevitable future outbreaks is apparent. We present a solution using existing sensors: Vital Screening Techniques (VIST). VIST involves scanning individual's multiple vitals within seconds using robust sensors. It provides the added layer of safety needed to move past COVID-19 pandemic. 

\section{Vitals Screening Techniques}
VIST encompasses any device that measures more than one independent vital( e.g., body temperature, heart rate, and blood oxygen level) of an individual. VIST devices usually have built in sensors in the form of thermal cameras and pulse oximeters. Most of the products display the readings on a user-friendly interface. A live video feed usually allows the user to adjust their positioning. VIST allows for the rapid mass screening of individuals to ensure a safe environment.

For external temperature sensing, thermal cameras are used to obtain targets' radiated temperature. The thermal camera should ideally be set up in a room temperature environment~\cite{FDA3,FDA4, FDA5}. Measurements should be made only at a fixed distance, with the subject directly facing the camera. The incorporation of a blackbody device allows for higher accuracy in the narrow range of normal human body temperature. The subject should acclimatize to room temperature for a few minutes prior to measurement. Per the FDA~\cite{FDA6}, it is recommended to measure one person at a time.  

In pulse oximetry, blood oxygen levels and heart rate are measured through the process of photoplethysmography (PPG). PPG is a non-invasive technology that uses a light source and a photodetector at the surface of the skin (finger-tip) to measure the volumetric variations of blood circulation with a resulting waveform. Pulse oximeters can incorporate either the transmissive or reflective mode. In the transmissive mode, the light sources and the photodiode are opposite to each other with the measurement site between them~\cite{lee2016reflectance}. In the reflective mode, the light sources and photodiode are on the same side, and light is reflected to the photodiode across the measurement site. The transmissive mode is not only the most commonly used method, it is the only clinically approved one, because of its high accuracy and stability. The clip-style of the pulse oximeter probe eliminates some of the errors due to finger movement. The only drawback of the clip style is that it is difficult to clean and cannot incorporate UV-C disinfection. 

A pulse oximeter used in mass screening is a high-touch surface with the potential for disease transmission. Ultraviolet (UV-C) sterilization is one method that may be incorporated to address this concern. UV-C’s effectiveness against different strains of viruses has long been established ~\cite{buonanno2020far}. Studies show that UV-C light at 267 and 279 nm was very effective at inactivating the Coronavirus~\cite{gerchman2020uv}.
Recent studies have shown the chance of transmission of SARS-CoV-2 (through inanimate surfaces) is less frequent than previously recognised ~\cite{mondelli2020low}. The improper exposure to UV-C radiation poses risks to human's skin and eyes. UV-C can be offered as an optional feature to alleviate concerns amongst the general public about touching this high-contact surface. 

 \begin{figure}[t!]
    \includegraphics[width=0.5\textwidth]{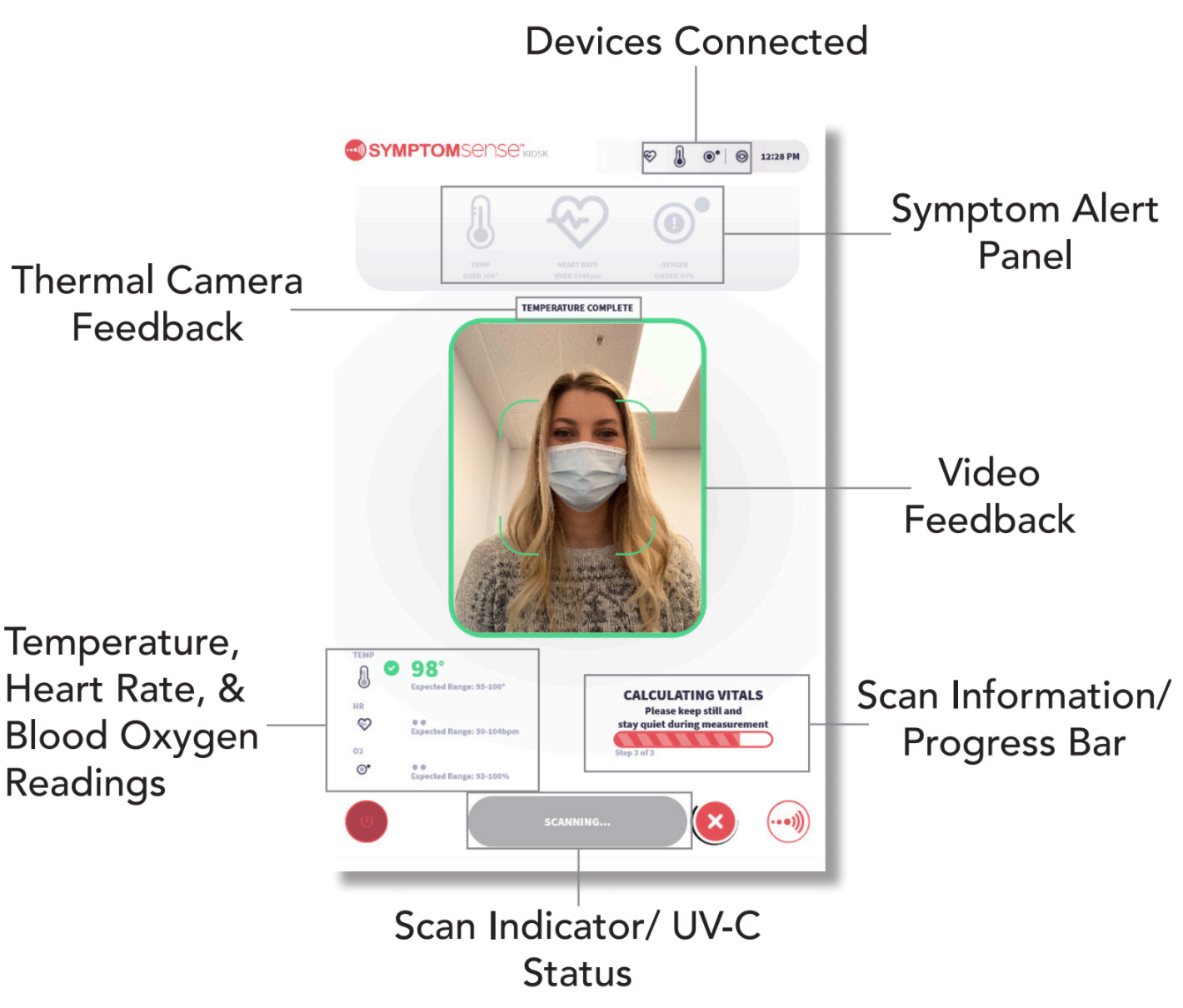}
\centering
\caption{ \textbf{Sample Interface from Symptomsense }}
\label{fig:testbed-dack}
\end{figure}

\section{Evaluation}
In this section, we evaluate Soter Technologies' sensors that are available for licensing, as well as full-integrated in products of various form factors (e.g., kiosk, handheld, gateway, etc.).
\subsection{Sensors Fidelity}

We compare Soter's sensors against other sensors on the market. The thermal sensor is compared against Braun thermometer ~\cite{Braun} and Famidoc~\cite{Famidoc}. Braun has FDA 510(K) Pre-market Notification, and Famidoc passed the EU standards for infrared thermometers accuracy. To the best of our knowledge, there is no existing remote thermal scanner that is FDA approved. The experiments were done over 20 different individuals with age ranges from 23 to 65 and various skin pigments. For each test, 10 samples were collected leading to total of 250 scans per each sensor. 

\textbf{Temperature Sensing.} Soter's thermal camera warms within 2 minutes after start-up. With an incorporated blackbody, ambient temperature does not affect readings. There is no drift or aging of the sensor due to the built-in blackbody technology. Multi-point temperature readings are made from different regions of the face. A unique weighted estimation algorithm is used adjust skin temperature to body temperature. Only one exposed region is necessary to obtain a reading, allowing for accurate temperature readings from individuals with head coverings, masks, etc.

We find that Famidoc has ±0.09°C (0.17°F) precision and Braun has ±0.12°C (0.21°F) precision. On average, Famidoc and Braun are about 0.40°C (0.72°F) from each others measurements. We find that Braun cannot measure people of lower than average body temperature (e.g. body temperature of 96°F or less). For lower body temperature individuals, we find that Famidoc performs better than Braun, however with degraded accuracy and precision performance. Soter's temperature sensor retains precision of ±0.28°C (0.51°F) and accuracy of ±0.15°C (0.27°F) compared to Braun. Compared to Famidoc, Soter's thermal sensor has an accuracy of ±0.18°C (0.33°F) Further, Soter's sensor is able to robustly detect those individual's with generally lower than average body temperature with high precision and accuracy relative to Braun and Famidoc. The average scan time of Soter's temperature sensor is 4.76 seconds. Further, we find that Soter's sensor yielded 100\% completed scans, unlike Braun, 97.5\% (Famidoc had 100\% completed scans). This shows that Soter's sensor can detect larger population than the Braun sensors. 
 
 \begin{figure}[t!]
    \includegraphics[width=0.5\textwidth]{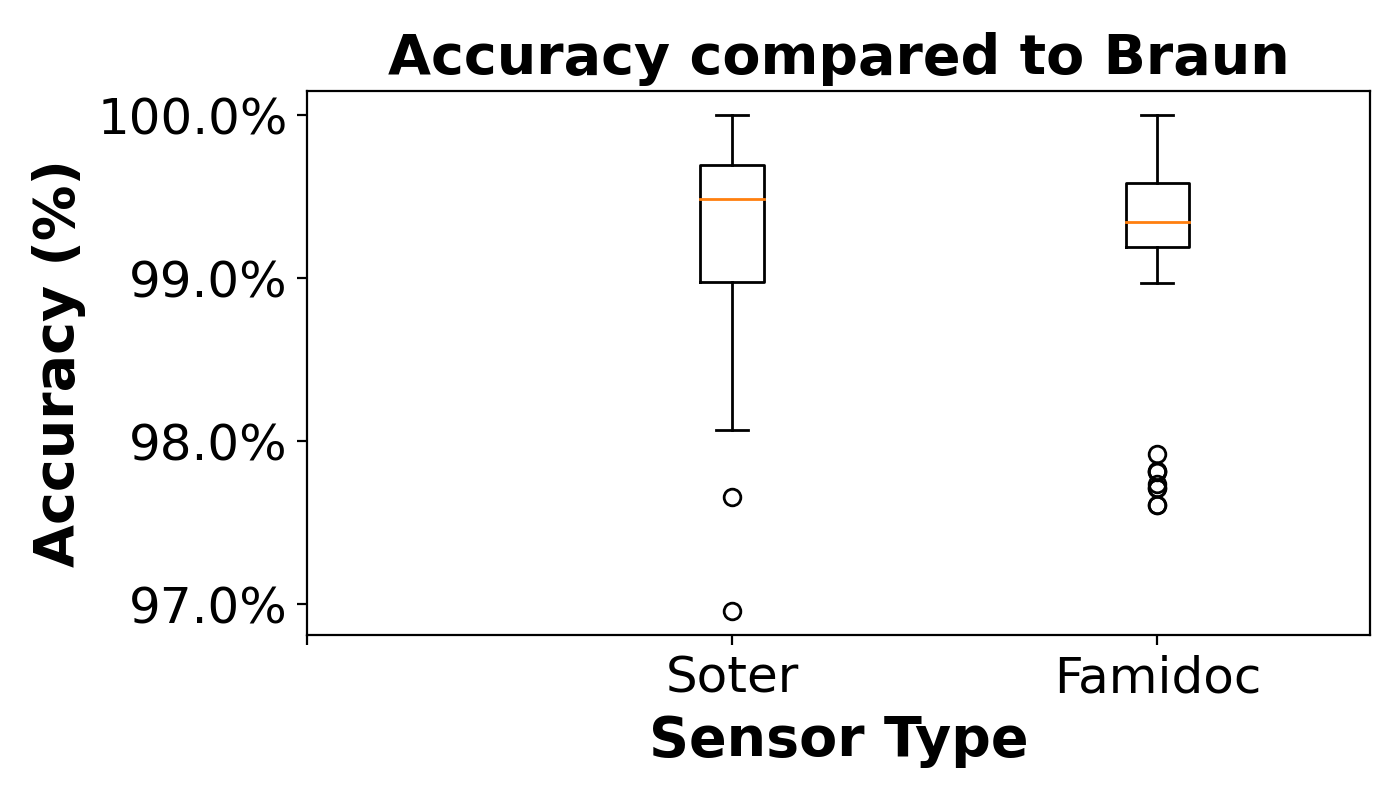}
\centering
\caption{ \textbf{Accuracy of Temperature Sensors }}
\label{fig:testbed-dack}
\end{figure}

\begin{table*}[]
\resizebox{1.05\textwidth}{!}{%
\begin{tabular}{|c|c|c|c|c|c|c|c|c|}
\hline
\multirow{2}{*}{Available} &
  \multirow{2}{*}{\checkmark} &
  \multirow{3}{*}{\textbf{\begin{tabular}[c]{@{}c@{}}SymptomSense\\  Kiosk\end{tabular}}} &
  \multirow{3}{*}{\textbf{\begin{tabular}[c]{@{}c@{}}SymptomSense \\ Gateway\end{tabular}}} &
  \multirow{3}{*}{\textbf{TAVIS}} &
  \multirow{3}{*}{\textbf{Olea Irvine}} &
  \multirow{3}{*}{\textbf{Olea Austin +}} &
  \multirow{3}{*}{\textbf{Ksubaka}} &
  \multirow{3}{*}{\textbf{Rapidscreen}} \\
                                         &                       &   &   &   &   &   &   &   \\ \cline{1-2}
Unavailable                              & X                     &   &   &   &   &   &   &   \\ \hline
\multirow{4}{*}{\textbf{Temperature}}    & Thermal Sensor        & \checkmark &\checkmark& \checkmark& \checkmark& \checkmark& \checkmark& \checkmark\\ \cline{2-9} 
                                         & Blackbody             & \checkmark& \checkmark& X & X & X & X & X \\ \cline{2-9} 
                                         & Live Video            & \checkmark& X & \checkmark& \checkmark& \checkmark& \checkmark& \checkmark\\ \cline{2-9} 
                                         & Facial Detection      & \checkmark& \checkmark& X & \checkmark& \checkmark& \checkmark& \checkmark\\ \hline
\multirow{2}{*}{\textbf{Pulse Oximeter}} & Blood Oxygen          & \checkmark& \checkmark& \checkmark& X & X & X & X \\ \cline{2-9} 
                                         & Heart Rate            & \checkmark& \checkmark& \checkmark & X & X & X & X \\ \hline
\multicolumn{1}{|l|}{\multirow{2}{*}{}}  & Configurations        & \checkmark& \checkmark &\checkmark & X & \checkmark & X & \checkmark \\ \cline{2-9} 
\multicolumn{1}{|l|}{}                   & Integration Potential &\checkmark&\checkmark&\checkmark & X & \checkmark & X & X \\ \hline
\end{tabular}%
}

\caption{{\textbf{In-depth comparison between existing products in the market ~\cite{Symptomsense,TAVIS, OLEA,Ksubaka,Rapidscreen}}. The table shows existing sensors that provide reliable temperature sensing by combining known techniques, thus enabling the highest sensing accuracy (e.g., facial detection, blackbody usage, etc.). We also show existing products in the market that support pulse oximetry sensor to provide oxygen levels and heart rate readings.}}
\label{tab:my-table}
\end{table*}

\textbf{Heart Rate and Blood Oxygen}
Soter's product line offers various fully-integrated pulse oximeters available in different form factors (open design, clip etc.). All of Soter's pulse oximeters use transmissive PPG technology. Here, we evaluate the off-the-shelf, medical-grade, FDA approved pulse oximeter (Nonin 3231)\footnote{Test subjects included those that traditionally may have difficulty obtaining pulse oximetry readings, including those with powder coated nails, poor perfusion, as well as cold hands.}. We find the average pulse oximetry scan time is 7.7 seconds from our testing. Nonin 3231 yielded 100\% successfully completed scans. This pulse oximeter is FDA approved and has undergone a clinical study with a stated accuracy of ±1.31\% between 70-100\% blood oxygen~\cite{Nonin3231}. Heart rate accuracy is ±3 digits between 20-250 bpm. 
 
The Nonin 3231 clip pulse oximeter measures a wide range of heart rates: 18 – 321 beats per minute. It can measure blood oxygen levels between 0-100\%. The ratio between the amplitude of the red light at 660nm and infrared light at 910nm wavelength is used to determine oxygen saturation. The pulse oximeter requires only 2 beats and uses an averaging algorithm to obtain an accurate heart rate. Longer scan times (with more readings taken) will result in more accurate pulse oximetry values.

\subsection{User Experience}

As an example of existing products' user interfaces, SymptomSense provides a user friendly interface (Figure 1) where users can read their results (either vital values or pass/fail) right after screening. Figure 1 shows a completed scan with temperature, heart rate and blood oxygen readings (on the left of the interface), and pulse oximetry scan complete (on the right). Once readings are obtained, visual indicators are used to indicate if the individual has passed the screening or not per the ranges of acceptable pass defined by the system operator. 

\subsection{Existing Products}
Table~\ref{tab:my-table} shows in-depth comparison between existing products that leverage both single and multi-vital sensing. Configurations include the ability to set  vital ranges, pass or fail vs  numeric readings. Integration features include an optional battery, printer, barcode system, etc.  


\section{Discussion}


\textbf{UV-C and standard medical pulse oximeter.} There is a shift to incorporate UV-C into pulse oximeters.
Soter's proprietary pulse oximeter with optional UV-C is currently in the final phase of development. This pulse oximeter utilizes a double-fail safe to automatically turn off when a finger is detected ensuring limited exposure to harmful UV-C rays. 275 nm of UV-C light (limited to the contact area) disinfects the surface for a 30 second duration after every use. To our knowledge, there is currently no FDA approved medical finger pulse oximeter with UV-C disinfection capability.

\textbf{Remote respiratory rate is in development.} There is no established technology that can remotely measure the respiratory rate with high confidence. Nonetheless, there is recent work with various technology to estimate and measure the respiratory rate (thermal sensing~\cite{chauvin2014contact}, mmwave ~\cite{alizadeh2019remote}). Given the current issues remote respiratory rate technology is facing (e.g., random individual vibrations, and better sensors required), it is not mature yet for the consumer market. However, to obtain a valid respiratory rate reading with high confidence, the individual has to get scanned for a long duration (e.g., > 10 seconds) to obtain a valid reading, since respiratory rate is much lower than heart rate (5-20).

\textbf{Screening best practices.} The FDA ~\cite{FDA1,FDA2, FDA3, FDA4, FDA5, FDA6} has provided proper guidelines on screening best practices (including thermal camera environment and calibration). Soter's sensors make the process of following the FDA procedure easier as they have self-calibrated blackbody based thermal system that shows final output to the user. Soter does not claim to do multi-person detection which has been shown to be ineffective. The Nonin 3231 pulse oximeter is already classified as a medical device and is FDA approved. 

\textbf{Further screening.} The entity using our kiosks may choose to do further screening/testing before rejecting the individual that our system flags ~\cite{FDA4}. We leave the guidelines of handling the user to the entity as it may differ and vary given location’s conditions. We recommend following CDC guidelines.

\textbf{HIPAA compliance.} Most of the screening systems do not store or retain any information about the user and their vitals, it only shows the vital results for a brief moment. Facial recognition technology is currently not implemented in the SymptomSense product-line, Thus, there is no need for HIPAA clearance, hence anonymized vital screening~\cite{gerke2020regulatory}.


\section{Acknowledgement}
We thank Asheik Hussain and Cary Chu for their invaluable feedback and continuous assistance of this work.

\section{Conclusion}
In this paper, we introduce VIST, vitals screening techniques. We cover state of the art vitals scanners that exist in the market, with in depth comparison between them. We further cover details about how the underlying sensing technologies work and their drawbacks.

{
\bibliographystyle{unsrt}
\bibliography{main.bib}
}
\end{document}